\documentclass[10pt,twoside]{article}
\usepackage{amis-Dixie}

\Volume{2}
\Issue{x}
\Year{2008}
\pages{xxx-xxx}
\FullTitle{Continuous variables approach to entanglement creation and processing}

\RunningTitle{Continuous variables approach to entanglement}

\RunningAuthor{Li-hui Sun et al.}

\Thanks{This work was supported by the National Natural Science Foundation of
China (60878004), the Ministry of Education under project NCET (Grant No. NCET-06-0671), 
and SRFDP (under Grant No. 200805110002)
}

\begin{document}

\amispage
\date{}


\author{\normalsize Li-hui Sun${}^1$, Gao-xiang Li${}^1$, and Zbigniew Ficek${}^2$\\
\small ${}^1$Department of Physics, Huazhong Normal University, Wuhan 430079, PR China\\
\small {\it Email Address: ~gaox@phy.ccnu.edu.cn}\\
\small ${}^2$The National Centre for Mathematics and Physics, KACST, P.O. Box 6086, Riyadh 11442, Saudi Arabia\\
\small {\it Email Address: ~zficek@kacst.edu.sa}\\
\footnotesize Received June 22, 200x; ~Revised March 21, 200x}
\maketitle

\thispagestyle{empty}


\begin{abstract}
\noindent We apply the continuous variable approach to study entangled dynamics of coupled harmonic oscillators interacting with a thermal reservoir and to a deterministic creation of entanglement in an atomic ensemble located inside a high-$Q$ ring cavity. In the case of harmonic oscillators, we show that a suitable unitary transformation of the position and momentum operators transforms the system to a set of independent harmonic oscillators with only one of them coupled to the reservoir. Working in the Wigner representation of the density operator, we find that the covariance matrix has a block diagonal form of smaller size matrices. This simple property allows to treat the problem to some extend analytically. We analyze the time evolution of an initial entanglement and find that the entanglement can persists for long times due to presence of constants of motion for the covariance matrix elements. In the case of an atomic ensemble located inside the cavity, the attention is focused on creation of one and two-mode continuous variable entangled states from the vacuum by applying laser pulses of a suitably adjusted amplitudes and phases. The pulses together with the cavity dissipation prepare the collective modes of the atomic ensemble in a desired entangled state.

\medskip




\end{abstract}

\section{Introduction}

Quantum information with continuous variables (CV) has attracted a great interest due to the simplicity in the generation, manipulation and detection of continuous variable states. Controlled dynamics and preservation of an initial entanglement encoded into  CV states are challenging problems in quantum information technologies~\cite{bp03} and have led to the development of different techniques for generation, manipulation and detection of CV multipartite entangled state~\cite{s1}. It has been shown that CV entangled states have many applications in quantum information processing such as quantum teleportation~\cite{s2}, dense coding~\cite{s3}, entanglement swapping~\cite{s4}, and quantum telecloning~\cite{s5}.

The continuous variables approach is mostly applied to creation and manipulation of entangled states produced with squeezed light and linear optics. A wide theoretical attention has been devoted to the application of the CV approach to the decoherence phenomena, in particular to the dissipative dynamics of open quantum systems, such as two coupled harmonic oscillators interacting with a dissipative reservoir. It has been analyzed with different approaches including the rotating-wave (RWA) and the  Born-Markov approximations that assume a weak coupling between a system and the reservoir.  A series of papers accounts these analysis for the case of two coupled harmonic oscillators interacting with a Markovian thermal reservoir and the work of Liu and Goan~\cite{lg07}, Hu {\it et al.}~\cite{paz92}, Maniscalco {\it et al.}~\cite{mo07} and H\"orhammer and B\"uttner~\cite{hb08} accounts for a non-Markovian thermal bosonic reservoirs. Detailed discussions and extensive reference lists devoted to the decoherence of two harmonic oscillators can be found in Refs.~\cite{b02,kl02,bf03,ok06,bl07,p02,sl04,dh04,pb04,bf06,az07,cb08,pa09,pr08}.
In all these studies a general conclusion made is that entanglement dynamics depends on the form of the reservoir and the non-Markovian nature of the reservoir preserves entanglement over a longer time. In this connection, we should mention work devoted to the variation of the entanglement sudden death phenomenon with the time scale of the evolution~\cite{Yu,ey07,ft06,cz08,mm09,jl09}.

Recently, a good deal of attention has been given to the problem of application of the~CV approach to creation and manipulation of entanglement in atomic systems~\cite{s12,s13,s14,lf10}. This is because a large collection of atoms can be efficiently coupled to quantum light and the existence of long atomic ground-state coherence lifetimes has been realized~\cite{s15}. Particularly interesting are the studies of the dynamics of collective atomic spin states of driven two-level atoms and coupled to a cavity field~\cite{s19}. It has been shown that the unconditional preparation of a two-mode squeezed state of effective bosonic modes can be realized in a pair of atomic ensembles interacting collectively with a two-mode optical cavity and laser fields~\cite{s20}. Another studies considered the deterministic creation of cluster states between atomic ensembles~\cite{lk09}.

In this paper, we illustrate applications of the CV approach to creation and processing of entanglement in two widely used systems: two coupled harmonic oscillators interacting with a non-Markovian thermal reservoir, and an ensemble of cold atoms located inside a a high-$Q$ ring cavity. In the first case of the coupled harmonic oscillators, we introduce an unitary transformation of the position and momentum operators and find that in the transformed basis the system is represented by a set of independent harmonic oscillators with only one of them coupled to the environment. This fact makes the problem remarkably simple that the relaxation properties of coupled harmonic oscillators follow the same pattern as a single harmonic oscillator. We work within the correlation matrix representation, also known as the covariance matrix, and find that the transformation of the position and momentum operators results in decoupled subsets of three and four equations of motion for the covariance matrix elements. This is particularly attractive for numerical analysis that it shorten the computations to the diagonalization or direct integration of small-size matrices. We then consider the time evolution of an initial entanglement encoded into the system of two coupled harmonic oscillators using the Gaussian continuous variable entangled states.

In the second case, we consider a practical scheme for creation of entangled states in an ensemble of cold atoms located inside a a high-$Q$ ring cavity. The scheme does not involve any external sources of squeezed light and networks of beam splitters used in the linear optics schemes~\cite{s11}. It involves an atomic ensemble driven by laser pulses and CV entangled states are created by specifically chosen Rabi frequencies and phases of the pulses. The atoms interact with the laser pulses and the cavity field in a highly nonresonant dispersive manner that involves ground states, not the excited states of the atoms. Therefore,  the process of creation of entangled states is not affected by the atomic spontaneous emission.

\section{Entangled dynamics of coupled harmonic oscillators}

We first consider entangled dynamics of a system composed of two mutually coupled identical harmonic oscillators of mass $M$ and frequency $\Omega$ that are simultaneously interacting with a thermal reservoir. The dynamics are determined in terms of Gaussian continuous variable entangled states, which are examples of multi-mode continuous variable entanglement states.
We introduce a two-mode unitary transformation of the position and momentum operators and show that in the case of a non-Markovian reservoir, the formalism allows us to obtain, to some extend, an analytical solution for the dynamics of the system, and the exact analytical solutions under the Markov approximation. Thus, they are much more suited to give a united picture of this complex system.

\subsection{Hamiltonian of the system}

The system is determined by the Hamiltonian, which in terms of the position $q_{i}$ and momentum $p_{i}$ operators  can be written~as
\begin{align}
H = H_{s}+H_{r}+V_{ss} +V_{sr} ,\label{e1}
\end{align}
where
\begin{align}
H_{s}=\sum^{2}_{i=1}\left(\frac{p^{2}_{i}}{2M} + {\frac{1}{2}}{M}{\Omega^{2}} q^{2}_{i}\right) \label{e2}
\end{align}
is the free Hamiltonian of the harmonic oscillators
\begin{align}
H_{r}=\sum_{n}\left(\frac{p^{2}_{n}}{2m_{n}}+{\frac{1}{2}}{m_{n}}
{\omega^{2}_{n}}{q^{2}_{n}}\right) \label{e3}
\end{align}
is the Hamiltonian of the reservoir, modeled as a set of independent oscillators of mass $m_{n}$ and frequency $\omega_{n}$ to which the systems' oscillators are coupled,
\begin{align}
V_{ss}=\lambda\sum^{2}_{i=1}\sum_{j>i}q_{i}q_{j} \label{e4}
\end{align}
represents the interaction between the oscillators with the coupling strength $\lambda$, and
\begin{align}
V_{sr}=\sum_{n}\sum_{i=1}^{2}{\lambda_{n}}{q_{n}}q_{i} \label{e5}
\end{align}
represents a bilinear interaction between the oscillators and the reservoir with the strength~$\lambda_{n}$.

The Hamiltonian~(\ref{e1}) is written in terms of $(q_{i},p_{i})$ operators, usually called as the bare basis. However, this is not a very convenient basis and, as we shall see, the dynamics of the systems' harmonic oscillators is most transparently discussed in terms of transformed position and momentum operators that are obtained by an $N$-mode unitary transformation of the position operators~\cite{pr08,l06}
\begin{align}
\tilde{q}_1 =\sqrt{\frac{1}{2}}\left(q_1-q_2\right) , \quad
\tilde{q}_2 = \sqrt{\frac{1}{2}}\left(q_1 + q_2\right) ,
\end{align}
and the same transformation of the momentum operators
\begin{align}
\tilde{p}_1 =\sqrt{\frac{1}{2}}\left(p_1-p_2\right) , \quad
\tilde{p}_2 = \sqrt{\frac{1}{2}}\left(p_1 + p_2\right) .
\end{align}
The transformed operators satisfy the well known position-momentum commutation relation, $[\tilde{q_i},\tilde{p}_j]=i\hbar\delta_{ij}$. We note that the transformations from $q_{i}$ and $p_{i}$ to $\tilde{q}_{i}$ and $\tilde{p}_{i}$  involve anti-symmetrical $(\tilde{q}_{1},\tilde{p}_{1})$ and symmetrical $(\tilde{q}_{2},\tilde{p}_{2})$ combinations of the position and the momentum operators, a close analog of the symmetric and antisymmetric multi-atom Dicke states~\cite{dic,ft02}.

In terms of the transformed operators, or equivalently in the transformed basis $(\tilde{q}_{i},\tilde{p}_{i})$, the Hamiltonian of the coupled oscillators $H_{s}+V_{s}$ takes the form
\begin{align}
\tilde{H}_{s} = H_{s}+V_{s}=\sum^{2}_{i=1}\left(\frac{\tilde{p}^{2}_{i}}{2M} + {\frac{1}{2}}{M}{\Omega^{2}_{i}} \tilde{q}^{2}_{i}\right) ,\label{e8}
\end{align}
where
\begin{align}
\Omega_1 \equiv \Omega_{F} =  \sqrt{\Omega^2-\frac{\lambda}{M}} ,\quad 
\Omega_2 =\sqrt{\Omega^2+\frac{\lambda}{M}} ,\label{e9}
\end{align}
and the interaction between the oscillators and the environment becomes
\begin{align}
V={\sqrt{2}\sum_{n}\lambda_{n}}{q_{n}}\tilde{q}_2 .\label{e10}
\end{align}

\begin{figure}[h]
\centering
\includegraphics[width=0.5\textwidth]{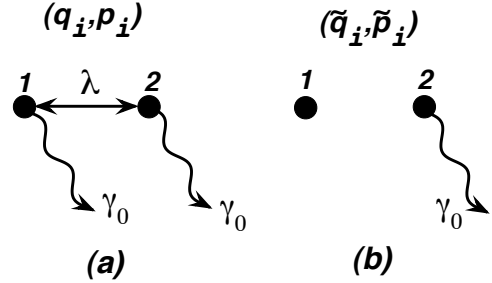}
\caption{A schematic diagram of a system of two coupled harmonic oscillators (a) before and (b) after the unitary transformation to new position and momentum basis.}
\label{fig:1a}
\end{figure}

One can see from Eqs.~(\ref{e8}) and (\ref{e10}) that in terms of the transformed operators, the system is represented by two {\it independent} oscillators with {\it only one} being coupled to the environment. The oscillator effectively coupled to the environment is that one corresponding to the symmetric combination of the position and momentum operators. In addition, the effective frequency $\Omega_{2}$ of the oscillator coupled to the environment differs from~$\Omega_{1}$, the decoupled oscillator. Figure~\ref{fig:1a} illustrates the result of the unitary transformation of the system to new position and momentum operators.

The oscillator effectively decoupled from the environment may be regarded as a decoherence-free oscillator. It should be stressed that the evolution of the oscillator is {\it not} free of the  decoherence. We shall demonstrate that the oscillator still can evolve in time that may lead to decoherence.

\subsection{The master equation}\label{sc3}

Let us now consider the evolution of the density operator of the system. We derive the master equation for the density operator and in the derivation we follow the standard method involving the Born approximation that corresponds to the second-order perturbative approach to the interaction between the oscillators and the environment, but we do not make the rotating-wave (RWA) and Markovian approximations. In the derivation, we assume that the system and the environment are uncorrelated at $t=0$, so that is the density operator of the total system factorizes $\rho_{T}(0) = \rho(0)\otimes\rho_r (0)$, where $\rho(0)$ is the density operator of the harmonic oscillators, that we shall call the "system" oscillators, and $\rho_{r}(0)$ is density operator of the reservoir. Further, we consider the reservoir is in a thermal state of temperature $T$ with the Boltzmann distribution of photons characterized by the mean occupation number
\begin{equation}
\bar{N}(\omega)=\frac{1}{\exp\left(\frac{\hbar\omega}{k_BT}\right) -1} .
\end{equation}

With these assumptions and after tracing over the reservoir modes, we arrive to the equation of motion for the reduced density matrix of the system of the form~\cite{lg07,paz92}
\begin{align}
\label{eq:masterequation}
\dot{\rho}(t) =& -\frac{i}{\hbar}[\tilde{H}_s+{\frac{1}{2}}M{\tilde{\Omega}^{2}_{2}(t)}{\tilde{q}_{2}^{2}},\rho(t)] - \frac{i}{\hbar}\gamma_2(t)[\tilde{q}_2,\{\tilde{p}_2,\rho(t) \}] \nonumber\\
&-D_2(t)[\tilde{q}_2,[\tilde{q}_2,\rho(t)]] -\frac{1}{\hbar}f_2(t)[\tilde{q}_2,[\tilde{p}_2,\rho(t)]] .
\end{align}
where $\tilde{\Omega}^{2}_{2}(t)$ represents a shift of the frequency of the oscillator due to the interaction with the environment, $\gamma_2(t)$ is the dissipation coefficient, and $D_2(t)$ and  $f_2(t)$ are diffusion coefficients. The parameters depend on the spectral density $J(\omega)$ of the reservoir modes. They also depend on time, which results from the non-Markovian nature of the coupling of the oscillators to the reservoir. The explicit forms of the parameters are given in Refs.~\cite{lg07,paz92}. As one could expect, the reservoir affects the evolution of only the oscillator 2 leaving the  oscillator 1 to evolve freely in time.

\subsection{Covariance matrix of two harmonic oscillators}

In order to analyze the entangled dynamics of the oscillators, we solve the master equation (\ref{eq:masterequation}). We apply the Wigner representation for the density operator. This requires a characteristic function, which can be written in terms of a covariance matrix, whose the elements are defined as
\begin{align}
V_{i,j} = {\rm Tr}\left\{ \frac{1}{2}\left(\Delta{X_i}\Delta{X_j}+\Delta{X_j}
\Delta{X_i}\right)\rho\right\} ,\label{e24}
\end{align}
with $\Delta{X_i} = {X_i}-\langle{X_i}\rangle$ and ${X} =({\tilde{q}}_1,{\tilde{p}}_1,{\tilde{q}}_2,{\tilde{p}}_2,\ldots, {\tilde{q}}_N,{\tilde{p}}_N)$.
Using the definition~(\ref{e24}) and the master equation~(\ref{eq:masterequation}) one can find the equations of motion for the covariance matrix elements that then can be solved for arbitrary initial conditions. The system of two harmonic oscillators is determined by a set of coupled linear differential equations for ten covariance matrix elements. In Refs.~\cite{lg07,paz92}, the set of differential equations was solved using numerical methods. In what follows, we  illustrate the advantage of working in the basis of the transformed position and momentum operators which will allow us to determine the covariance matrix elements in an effectively easy way requiring to solve separate sets of equations composed of three and four coupled differential equations.

The set of inhomogeneous differential equations for the covariance matrix elements can be written in a matrix form as
\begin{align}
\dot{\vec{V}}(t) = {\bf A}(t)\vec{V}(t) +\hbar \vec{F}(t)  ,\label{e43}
\end{align}
where we put the covariance matrix elements in a specific order in the column vector
\begin{align}
\vec{V}(t) ={\rm col}(V_{11},V_{12},V_{22},V_{13},V_{14},V_{23},V_{24},V_{33},V_{34},V_{44}) .
\end{align}
This specific order of the elements $V_{ij}$ leads to a block diagonal form of the matrix of the coefficients ${\bf A}(t)$ with two blocks each involving three equations and one block involving four equations.
The column vector of the inhomogeneous terms is of the form
\begin{align}
\vec{F}(t) ={\rm col}(0,0,0,0,0,0,0,0,- f_2(t),2\hbar D_2(t)) .
\end{align}
The remaining elements are found from the symmetry property of the covariance matrix,~$V_{ij}=V_{ji}$.

From the set of three coupled equations of motion for the elements $V_{11},V_{12},V_{22}$, one can find that there is a liner combination 
\begin{align}
V^{+}_{11} = M\Omega_{F}^{2}V_{11} +\frac{1}{M}V_{22} ,
\end{align}
which equation of motion is $\dot{V}^{+}_{11} =0$ and the remaining elements form a set of two coupled equations
\begin{align}
\dot{V}^{-}_{11} = 4\Omega_{F}^{2}V_{12} ,\quad 
\dot{V}_{12} = -V^{-}_{11} ,\label{e31}
\end{align}
where $V^{-}_{11}= M\Omega_{F}^{2}V_{11} -(1/M)V_{22}$.

The property of $\dot{V}_{11}^{+}=0$ indicates that the linear combination $V_{11}^{+}$ is a {\it constant of motion}, i.e. $V_{11}^{+}(t) =V_{11}^{+}(0)$. In other words, $V_{11}^{+}(t)$ does not change in time and retains its initial value for all times.
Physically, if initially the system was prepared in a state such that $V_{11}^{+}(0)\neq 0$ and with the other elements of the covariance matrix equal to zero, it would remain in that state for all times. For example, if the initial state is an entangled state, the initial entanglement of the system will remain constant in time. Therefore, the subspace composed of the $V_{11}^{+}(t)$ element can be regarded as a {\it decoherence-free subspace}.

The remaining matrix elements $V^{-}_{11}$ and $V_{12}$ can undergo a temporal evolution.  Since there is no damping involved in the equations of motion  (\ref{e31}), the solution would lead to the matrix elements continuously oscillating in time. It is easy to see, the solution of Eq.~(\ref{e31}) has a simple form
\begin{align}
V^{-}_{11}(t) &= V^{-}_{11}(0)\!\cos(2\Omega_{F}t) - 2\Omega_{F}V_{12}(0)\sin(2\Omega_{F}t) ,\nonumber\\
V_{12}(t) &=  V_{12}(0)\!\cos(2\Omega_{F}t) - \frac{V^{-}_{11}(0)}{4\Omega_{F}}\sin(2\Omega_{F}t) ,\label{e32a}
\end{align}
from which we see the matrix elements continuously oscillate in time with frequency $2\Omega_{F}$. This indicates that the system will never reach a stationary time-independent state unless $V^{-}_{11}(0)=V_{12}(0)=0$. As we shall see, this feature will result in a continuous in time entanglement in the system.

\subsection{Initial two-mode squeezed state}

We have already seen that due to the presence of the constants of motion in the evolution of the covariance matrix elements, the dynamics of the system, even after a long time, may strongly depend on the initial state. Since we are interested in the evolution of an initial two-mode entangled state and it is well known that two-mode squeezed states are examples of entangled states, we consider an initial squeezed vacuum state. We also demonstrate that with the specific initial state, the problem of treating the dynamics of two harmonic oscillators simplifies to analysis of the properties of only those constants of motion and the matrices which involve only the diagonal elements of the covariance matrix.
As an initial state, we consider the following two-mode continuous variable squeezed state
\begin{align}
|\psi_0\rangle= {\rm e}^{r\left(b_1 b_2 - b_{1}^{\dagger}b_{2}^{\dagger}\right)}|0_{b_1}\rangle|0_{b_2}\rangle ,\label{e32}
\end{align}
where $r$ is the squeezing parameter and the kets $|0_{b_1}\rangle$ and $|0_{b_2}\rangle$ represent the state with zero photons in the modes 1 and 2, respectively.

With the initial state (\ref{e32}), we easily find that non-zero initial covariance matrix elements are
\begin{align}
V^{\pm}_{11}(0) = \frac{1}{2}\left(M\Omega_{F}^{2}{\rm e}^{-2r} \pm \frac{1}{M}{\rm e}^{2r}\right) ,\ 
V^{\pm}_{33}(0) = \frac{1}{2}\left(M\Omega_{F}^{2}{\rm e}^{2r} \pm \frac{1}{M}{\rm e}^{-2r}\right) .
\end{align}
Note that the elements $V^{\pm}_{11}(0)$ and $V^{\pm}_{33}(0)$ have opposite behavior with the squeezing correlations $r$.
The frequency dependent part of $V^{\pm}_{11}(0)$ is squeezed while the other part is enhanced by the correlations. On the other hand, the frequency dependent part of $V^{\pm}_{33}(0)$ is enhanced and the other part squeezed by the correlations. This is a crucial difference that will have a significant effect on the evolution of an entanglement injected into the system.

\subsection{Time evolution of an initial entanglement}\label{sc6}

We now perform numerical analysis of time evolution of entanglement between two harmonic oscillators simultaneously coupled to a reservoir. We will illustrate the advantage of working in the transformed basis to obtain a simple interpretation of the results. In particular, to understand short time non-Markovian dynamics of entanglement and to provide conditions for optimal and stable long time entanglement. We adopt the negative partial transpose criterion that is known as the necessary and sufficient condition for entanglement of a two-mode Gaussian state~\cite{SeparabilityCriterion_Simon00,dg00}.  We will use this criterion to quantify the amount of entanglement in the system, and will denote it by a parameter $\eta_-$. In all examples considered here, we assume that the oscillators interact with a reservoir of temperature~$T=10\hbar\Omega/k_{B}$ and were initially in the two-mode squeezed state $|\psi_0\rangle$. Moreover, we assume a Gaussian-type spectral density for the reservoir modes
\begin{align}
J(\omega)=\frac{2\gamma_0\omega M}{\pi}\left(\frac{\omega}{\Lambda}\right)^{n-1}
{\rm e}^{-\omega^2/\Lambda^2} ,
\end{align}
where $M$ is the mass of system, $\Lambda$ is cut-off frequency that represents the highest frequency in the reservoir, $\gamma_0$ is proportional to the coupling strength between the  oscillators and the environment and $n$ determines the type of the reservoir. We choose $n=1$, the so-called Ohmic reservoir.
\begin{figure}[h]
\centering
\includegraphics[width=0.6\textwidth]{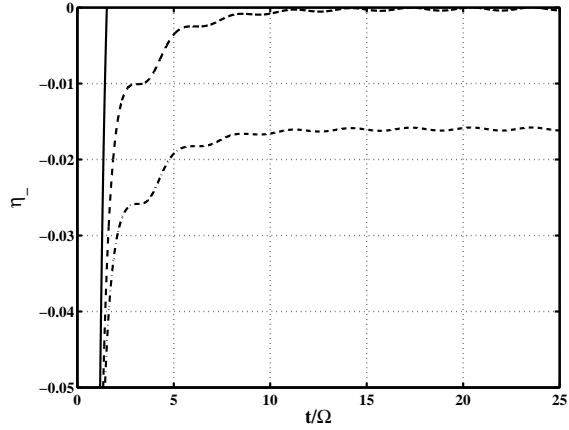}
\caption{Time evolution of the negativity $\eta_-$ for $\gamma_0=0.1, \Lambda=100, \lambda=0$ and different~$r$: $r=1.0$ (solid line), $r=1.498$ (dashed line), $r=2.0$ (dashed-dotted line).}
\label{lifig2}
\end{figure}

We first consider the case of mutually independent oscillators with $\lambda=0$, but coupled to a  thermal reservoir. The interaction with the reservoir is usually recognized as an irreversible loss of entanglement and information encoded in the internal states of the system and thus is regarded as the main obstacle in practical implementation of entanglement.
Figure~\ref{lifig2} shows the negativity as a function of time for the initial two-mode squeezed state with different degree of squeezing $r$. First of all, we note that there is a threshold value for the degree of squeezing $r$ at which a continuous in time entanglement occurs~\cite{pb04}. The threshold that corresponds to entanglement undergoing the phenomenon of sudden death, occurs at $r=1.498$. 

The presence of the threshold value for $r$ at which continuous in time entanglement occurs has a simple interpretation in terms of the covariant matrix elements. Consider the threshold in the long time limit in which we may consider the evolution under the Markov approximation, but retaining the non-RWA terms. Under this approximation, we can put $\gamma_{2}(t)\rightarrow \gamma_{0}$ which then allows us to obtain a simple analytical solution for the threshold condition for entanglement. It is easy to show that the threshold for two mode entanglement occurs at
\begin{align}
V_{11}(t)V_{44}(t) = 1 ,\label{e60}
\end{align}
so that the two modes are entangled when $V_{11}(t)V_{44}(t) <1$, otherwise are separable.

Note that the covariance matrix element $V_{11}(t)$ is associated with the relaxation free modes whereas the element $V_{44}(t)$ is associated with the mode that is coupled to the reservoir and thus undergoes the damping process. Under the Markov approximation, we find from Eq.~(\ref{e43}) that in the long time limit of $t\gg \gamma_{2}^{-1}$, the element $V_{44}(t)$ reaches the stationary value equal to the level of the thermal fluctuations
\begin{align}
V_{44}(t)\rightarrow 2{\bar N}+1 ,\label{e61}
\end{align}
whereas $V_{11}(t)$ retains its time dependent behavior which depends on the initial values
\begin{align}
V_{11}(t) = V_{11}(0)\cos^{2}\Omega_{F}t
+\frac{V_{22}(0)}{M^{2}}\left(\frac{\sin\Omega_{F}t}{\Omega_{F}t}\right)^{2} t^{2} .\label{e62}
\end{align}
We point out that the dependence on the initial values of the long time behavior of $V_{11}(t)$ is due to the presence of the constant of motion $V_{11}^{+}$.

Averaging Eq.~(\ref{e62}) over a long period of oscillations, the threshold condition (\ref{e60}) simplifies to
\begin{align}
2V_{11}(0)\left(2{\bar N}+1\right) = 1.\label{e63}
\end{align}
We see that the threshold behavior of entanglement depends on the initial value of the covariance matrix element $V_{11}(0)$. In other words, the entanglement behavior can be controlled by the suitable choosing of the initial state. For example, with the initial state (\ref{e32}), $V_{11}(0)=\exp(-2r)/2$, and then we find from Eq.~(\ref{e63}) that continuous entanglement occurs for the degree of squeezing
\begin{align}
r \geq \frac{1}{2}\ln\left(2{\bar N}+1\right) .\label{e64}
\end{align}
With the parameter value $T=10\hbar\Omega/k_{B}$, we find that the threshold value for $r$ equals to~$1.498$ that is the same found numerically and plot in Fig.~\ref{lifig2}. We should point out here that the same condition for the threshold value of $r$ has been found under the RWA approximation~\cite{pb04}. Thus, we may conclude that the threshold value for continuous entanglement is not sensitive to the RWA approximation.
\begin{figure}[h]
\centering
\includegraphics[width=0.6\textwidth]{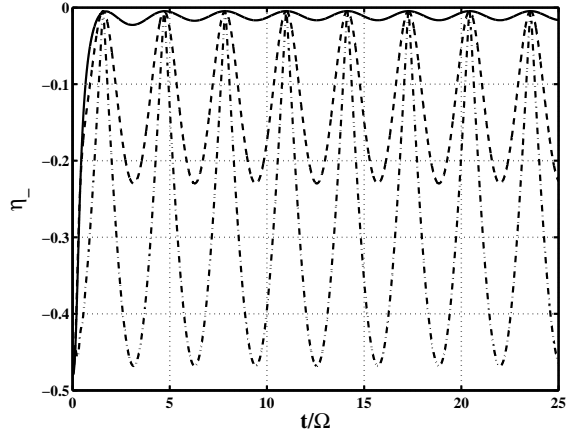}
\caption{Time evolution of the negativity $\eta_-$ for $\lambda=0, r=1.6, \gamma_{0}=1.0$ and different values of the cut-off frequency $\Lambda$ :$\Lambda=200$ (solid line), $\Lambda=500$ (dashed line), $\Lambda=800$ (dashed-dotted line).}
\label{lifig3}
\end{figure}

The preservation of an initial entanglement over a long time is usually related to the non-Markovian nature of the reservoir. We stress that the continuous in time oscillations are related to the Markovian rather than to the non-Markovian nature of the reservoir as the matrix elements $V_{11}(t)$ and $V_{22}(t)$ determine dynamics of the oscillator that is not coupled to the reservoir. To show this more qualitatively, we plot in Fig.~\ref{lifig3} the time evolution of an initial entanglement for different $\Lambda$, corresponding to the bandwith of the reservoir. It is evident from the figure that the amplitude of the oscillations increases with increasing~$\Lambda$ indicating that the oscillations are associated with broadband rather than narrow-band nature of the reservoir.

\begin{figure}[h]
\centering
\includegraphics[width=0.6\textwidth]{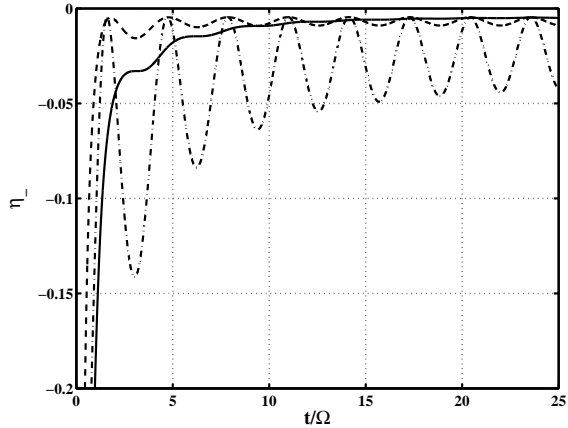}
\caption{Time evolution of the negativity $\eta_-$ for $\Lambda=100, \lambda=0, r=1.6$ and different values of the relaxation rate $\gamma_{0}$: $\gamma_{0}=0.05$ (solid line), $\gamma_{0}=1.0$ (dashed line), $\gamma_{0}=5.0$ (dashed-dotted line).}
\label{lifig4}
\end{figure}

It is also interesting to discuss the dependence of the long time entanglement on the relaxation rate $\gamma_{0}$. An example of this feature is shown in Fig.~\ref{lifig4}. It is interesting to note that under the relaxation the entanglement oscillates in time and the amplitude of the oscillations increases with increasing $\gamma_{0}$ leading to a better entanglement when the oscillators are strongly damped. It is a surprising result as one could expect that entanglement should decrease with increasing $\gamma_{0}$. Again, a straightforward interpretation of this effect can be gained from a qualitative inspection of the properties of the transformed covariance matrix.

It is easily verified that in the limit of vanishing damping, $\gamma_{2}(t)\rightarrow 0$ and $\lambda =0$, the equations of motion of the matrix elements $V_{33},V_{34},V_{44}$ reduce to that of the matrix elements $V_{11},V_{12},V_{22}$. One could argue that in this limit the covariance matrix elements $V_{33},V_{34},V_{44}$ coincidence with the elements $V_{11},V_{12},V_{22}$. Of course, their time evolution is determined by the same equations, but there is a subtle difference in their initial values. For example, the initial values of the linear combinations $V^{\pm}_{11}$ are
\begin{align}
V^{\pm}_{11}(0) = \frac{1}{2}\left(M\Omega_{F}^{2}{\rm e}^{-2r} \pm \frac{1}{M}{\rm e}^{2r}\right) ,
\end{align}
whereas
\begin{align}
V^{\pm}_{33}(0) = \frac{1}{2}\left(M\Omega_{F}^{2}{\rm e}^{2r} \pm \frac{1}{M}{\rm e}^{-2r}\right) .
\end{align}
The initial elements are significantly different that what appears as a squeezed component in $V^{\pm}_{11}(0)$, the counterpart in $V^{\pm}_{33}(0)$ appears as an anti-squeezed component. This is a crucial difference that has a significant effect on the evolution of an entanglement. These two contributions cancel each other that results in no oscillations in the entanglement evolution when $\gamma_{0}\ll 1$. On the other hand, for large~$\gamma_{0}$ the covariance matrix elements $V_{33},V_{34},V_{44}$ are rapidly damped to their stationary values leaving the elements $V_{11},V_{12},V_{22}$ continuously oscillating in time. These oscillations lead to continuous oscillations of the entanglement seen in Fig.~\ref{lifig4}.

The same arguments apply for the presence of the oscillations in the limit of a large $\Lambda$, seen in Fig.~\ref{lifig3}. Using similar arguments, one can finds that for large $\Lambda$, the covariance matrix elements $V_{33},V_{34},V_{44}$ are rapidly damped to their stationary values leaving the elements $V_{11},V_{12},V_{22}$ continuously oscillating in time.
\begin{figure}[h]
\centering
\includegraphics[width=0.6\textwidth]{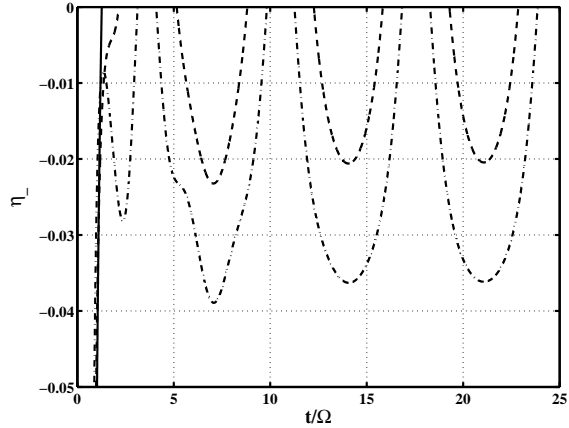}
\caption{Time evolution of the negativity $\eta_-$  for $\gamma_0=0.1, \Lambda=100,  \lambda=0.8$ and different $r$: $r=1.0$ (solid line), $r=1.498$ (dashed line), $r=2.0$ (dashed-dotted line).}
\label{lifig5}
\end{figure}

Figure~\ref{lifig5} shows the evolution of entanglement when the oscillators are coupled to each other. In this case there is no continuous stationary entanglement. Thus, the interaction between the oscillators has a destructive effect on the stationary entanglement. However, for a large squeezing, entanglement re-appears in some discrete periods of time, exhibiting periodic sudden death and revival of entanglement. In other words, the threshold behavior of entanglement is a periodic function of time.

As before, this feature has a simple interpretation in terms of the covariance matrix elements. According to Eq.~(\ref{e60}), for a given temperature the threshold value for entanglement depends on the covariance matrix element~$V_{11}(t)$ which, on the other hand, depends on~$\lambda$ through the frequency parameter $\Omega_{F}$. We see from Eq.~(\ref{e9}) that the frequency $\Omega_{F}$ decreases with increasing~$\lambda$. Thus, according to Eq.~(\ref{e62}) for interacting oscillators the matrix element~$V_{11}(t)$ oscillates slowly in time. In this case, the averaging over the oscillations is not justified and thus the threshold condition for entanglement is the oscillating function of time even in a long time regime.

\section{Creation of CV entangled states in an atomic ensemble}

\begin{figure}[h]
\centering
\includegraphics[width=0.4\textwidth]{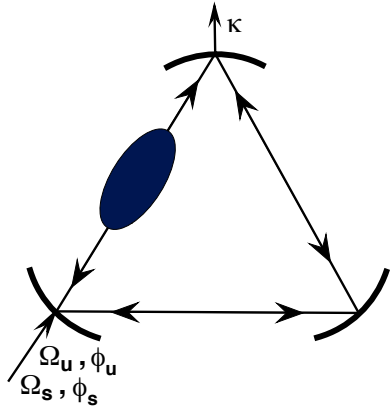}
\caption{A schematic diagram of a two-mode ring cavity containing an ensemble of cold atoms trapped along the cavity axis. The cavity modes are damped with the same rate $\kappa$. The driving laser fields of the Rabi frequencies $\Omega_{u},\Omega_{s}$ and phases $\phi_{u},\phi_{s}$ are injected through the cavity mirrors and co-propagate with one of the cavity modes.}
\label{lifig6}
\end{figure}

In this section, we consider a completely different system that addresses the issue of the creation of CV entangled states in an atomic system. The system we consider is composed of an atomic ensemble located inside a high-$Q$ ring cavity with two mutually counter-propagating modes, as illustrated in Fig.~\ref{lifig6}. 
In this two-mode configuration, the photon redistribution can occur between the counter-propagating modes which, as we shall see, may result in multi-mode squeezing. The procedure is associated with the collective dynamics of the atomic ensembles subjected to driving lasers of a suitably adjusted amplitudes and phases.

Recently, Parkins {\it et al.}~\cite{s20} have proposed a scheme for preparing single and two-mode squeezed states in atomic ensembles located inside a single mode high-$Q$ ring cavity. The scheme is based on a suitable driving of the atomic ensembles with two external laser fields that prepares the atoms in a pure squeezed (entangled) state. Similar schemes have been proposed to realize an effective Dicke model operating in the phase transition regime~\cite{de07}, to create a stationary subradiant state in an ultracold atomic gas~\cite{cb09}. This approach has recently been proposed as a practical scheme to prepare trapped and cooled ions in pure entangled vibrational states~\cite{li06} and to prepare four ensembles of hot atoms in pure entangled cluster states~\cite{lk09}.
The scheme proposed by Parkins {\it et al.}~\cite{s20} assumes that atomic ensembles are effectively coupled to a single or two modes co-propagating with the driving fields.  In fact, a practical ring cavity is composed of two degenerate mutually counter-propagating modes that can simultaneously couple to the atomic ensembles with the same coupling strengths~\cite{krb03,na03,kl06}. Here, we explore the role of the counter propagating mode in the creation of one and two-mode entangled states in a single atomic ensemble.

\subsection{Hamiltonian of the system}

We consider a system composed of a single atomic ensemble located along the axis of a high-$Q$ ring cavity, as illustrated in Fig.~\ref{lifig6}. The cavity is composed of three mirrors that create two mutually counter-propagating modes, called {\it clockwise} and {\it anti-clockwise} modes, to which the atoms are equally coupled. The cavity modes are degenerate in frequency, i.e.  $\omega_{+}=\omega_{-}=\omega_{c}$ and are damped with the same rate~$\kappa$.
The atomic ensemble contains a large number of identical four-level atoms, each composed of two ground states, $|0_{j}\rangle$, $|1_{j}\rangle$, and two excited states $|u_{j}\rangle$, $|s_{j}\rangle$, where the subscript $j$ labels atoms in the ensemble. The ground state $|0_{j}\rangle$ of energy $E_{0}=0$  is coupled to the excited state $|s_{j}\rangle$ by a laser field of the Rabi frequency $\Omega_{s}$ and frequency $\omega_{Ls}$ that is detuned from the atomic transition frequency by $\Delta_{s} = (\omega_{s}-\omega_{Ls})$, where $\omega_{s}=E_{s}/\hbar$ and $E_{s}$ is the energy of the state $|s_{j}\rangle$. Similarly, the ground state $|1_{j}\rangle$ of energy $E_{1}=\hbar\omega_{1}$ is coupled to the excited state $|u_{j}\rangle$ of energy $E_{u}=\hbar\omega_{u}$ by an another laser field with the Rabi frequency $\Omega_{u}$, and the angular frequency $\omega_{Lu}$ that is detuned from the atomic transition $|1_{j}\rangle\rightarrow|u_{j}\rangle$ by $\Delta_{u}= (\omega_{u}-\omega_{1}-\omega_{Lu})$. The frequencies $\omega_{Lu}$ and $\omega_{Ls}$ of the laser fields are matched close to the cavity frequency, so that the wave numbers of the laser fields $k_{u}$ and $k_{s}$ are approximated by $k_u\approx k_s\approx k$.

The Hamiltonian of the system, in the rotating-wave approximation and  in the interaction picture, can be written as
\begin{eqnarray}
H = H_{0}+H_{AL}+H_{AC} ,\label{12.1}
\end{eqnarray}
where
\begin{eqnarray}
H_{0} &=& \hbar\sum\limits_{j=1}^N  \left(\omega_u|u_{j}\rangle\langle
u_{j}| +\omega_s|s_{j}\rangle\langle s_{j}|+\omega_1|1_{j}\rangle\langle
1_{j}|\right)\nonumber\\
&+&  \hbar\omega_c (a_+^\dag a_{+} +a_{-}^{\dag} a_{-}) \label{12.2}
\end{eqnarray}
is the free Hamiltonian of the atoms and the cavity modes,
\begin{eqnarray}
 H_{AL} &=&  \frac{1}{2}\hbar\sum\limits_{j=1}^N \left\{\Omega_{u}{\rm e}^{-i[(\omega_{Lu} t+\phi_{u})-k_{u}x_{j}]}|u_{j}\rangle\langle 1_{j}|\right. \nonumber\\
&+&\left. \Omega_{s} {\rm e}^{-i[(\omega_{Ls} t+\phi_{s})-k_{s}x_{j}]}
|s_{j}\rangle\langle 0_{j}| + {\rm H.c.}\right\} \label{12.3}
\end{eqnarray}
is the interaction Hamiltonian between the atoms and the driving laser fields, and
\begin{eqnarray}
 H_{AC} &=& \hbar\sum\limits_{j=1}^N\left\{g_{u}^{(+)}{\rm e}^{ikx_{j}}|u_{j}\rangle\langle0_{j}|a_{+}
+ g_{u}^{(-)}{\rm e}^{-ikx_{j}}|u_{j}\rangle\langle0_{j}|a_{-}\right. \nonumber\\
&+&\left. g_{s}^{(+)}{\rm e}^{ikx_{j}}|s_{j}\rangle\langle1_{j}|a_{+}
+ g_{s}^{(-)}{\rm e}^{-ikx_{j}}|s_{j}\rangle\langle1_{j}|a_{-} +{\rm H.c.}\right\}  \label{12.4}
\end{eqnarray}
is the interaction Hamiltonian between the atoms and the cavity modes. The notation used here should be interpreted as follows. The operators $a_{\pm}$ and $a^{\dagger}_{\pm}$ are the annihilation and creation operators of the two counter-propagating modes of the cavity; clockwise and anti-clockwise propagating modes, and $k$ is their wave number. The parameters
$g_{u}^{(\pm)}$ and $g_{s}^{(\pm)}$ are the coupling constants of the atomic transitions to the cavity modes, $\Omega_{u}$ and $\Omega_{s}$ are Rabi frequencies of the driving laser fields, and $\phi_{u},\phi_{s}$ are their phases. We have set the energy of the ground state $|0_{j}\rangle$ equal to zero and have denoted the energies of the excited atomic levels by $\hbar\omega_{i} \, (i=1,u,s)$.

We now introduce detunings of the laser fields from the atomic transition frequencies
\begin{eqnarray}
\Delta_u = (\omega_u-\omega_{1}) - \omega_{Lu} ,\quad \Delta_s=\omega_s-\omega_{Ls} ,\label{12.7}
\end{eqnarray}
and assume that the laser frequencies satisfy the resonance condition
$\omega_{Ls}-\omega_{Lu}=2\omega_1$. Next, we make few standard approximations on the Hamiltonian to eliminate the excited states to neglect atomic spontaneous emission and to obtain an effective two-level Raman-coupled Hamiltonian.
In order to eliminate spontaneous scattering of photons to modes other than the privileged cavity mode, we assume that the detunings are much larger than the Rabi frequencies, cavity coupling constants and the atomic spontaneous emission rates,~i.e.
\begin{eqnarray}
\Delta_u, \Delta_s\gg  g_{s}^{(\pm)},g_{u}^{(\pm)}, \Omega_{s}, \Omega_{u}, \gamma_{s},\gamma_{u} , \label{12.8}
\end{eqnarray}
where $\gamma_{s}$ and $\gamma_{u}$ are the total spontaneous emission rates from the states $|s_{j}\rangle$ and $|u_{j}\rangle$, respectively.
The assumption about large detunings allows us to perform the standard adiabatic elimination of the atomic excited states, $|s_{j}\rangle$, $|u_{j}\rangle$, and obtain an effective two-level Hamiltonian involving only the ground states of the atoms
\begin{eqnarray}
H_{e} &=& \hbar\left[\delta_{c} + \frac{N}{2}\left(
\frac{g_{u}^{2}}{\Delta_{u}} + \frac{g_{s}^{2}}{\Delta_{s}}\right) + \left( \frac{g_{u}^{2}}{\Delta_{u}} -\frac{g_{s}^{2}}{\Delta_{s}}\right) J_{z}\right]\left(a_+^\dag a_+ +a_-^\dag a_{-}\right) \nonumber\\
&+& \frac{\hbar}{\sqrt{N}}\left\{\,
\beta_{u}\left(J_{0k}a_+^\dag +J_{2k}a_-^\dag \right)
+ \beta_{s}\left(J^{\dagger}_{0k} a_+^\dag + J^{\dagger}_{-2k}a_-^\dag\right) + {\rm H.c.}\right\} ,\label{12.9}
\end{eqnarray}
where we have assumed that the coupling constants $g_{u}\equiv g_{u}^{(\pm)}$ and~$g_{s}\equiv g_{s}^{(\pm)}$ are the same for both modes, $\delta_{c} =\omega_{c}-(\omega_{Ls}-\omega_{1})$ is the detuning of the cavity frequency from the Raman coupling resonance,
\begin{eqnarray}
J_{z} = \frac{1}{2}\sum\limits_{j=1}^N\left(|1_{j}\rangle\langle 1_{j}|-|0_{j}\rangle\langle 0_{j}|\right) ,\quad  J_{mk} = \sum\limits_{j=1}^{N}|0_{j}\rangle\langle 1_{j}|{\rm e}^{imkx_{j}} \label{12.10}
\end{eqnarray}
are position dependent collective atomic operators, and
\begin{eqnarray}
\beta_{u}=\sqrt{N}\frac{\Omega_{u}g_{u}}{2\Delta_{u}}{\rm e}^{-i\phi_{u}} ,\quad
\beta_{s}=\sqrt{N}\frac{\Omega_{s}g_{s}}{2\Delta_{s}}{\rm e}^{-i\phi_{s}} \label{12.11}
\end{eqnarray}
are the coupling strengths of the effective two-level system to the cavity modes.

The first line in Eq.~(\ref{12.9}) represents the free energy of the atomic ensemble and the intensity dependent (Stark) shift of the atomic energy levels. The second line represent the interaction between the cavity field and the atomic ensemble. The three collective atomic operators, $J_{0k}, J_{2k}$ and $J_{-2k}$, which appear in Eq.~(\ref{12.9}), arise naturally for the position dependent atomic transition operators and appear in a cavity with two mutually counter-propagating modes. In the case of a single-mode cavity, the Hamiltonian involves only the~$J_{0k}$ operator~\cite{s20,de07,cb09}. 

To avoid unessential complexity due to the presence of the free energy and the Stark shift terms in the Hamiltonian (\ref{12.9}), we will work with a simplified version of the Hamiltonian by choosing the frequencies of the driving and the cavity fields such that
\begin{eqnarray}
\delta_{c} + \frac{N}{2}\left(\frac{g_{u}^{2}}{\Delta_{u}} + \frac{g_{s}^{2}}{\Delta_{s}}\right) = 0 \quad {\rm and}\quad \frac{g_{u}^{2}}{\Delta_{u}} = \frac{g_{s}^{2}}{\Delta_{s}} . \label{12.12}
\end{eqnarray}
With this choice of the parameters, the Hamiltonian~(\ref{12.9}) reduces to
\begin{eqnarray}
H_{e} = \frac{\hbar}{\sqrt{N}}\left\{\,
\beta_{u}\left(J_{0k}a_+^\dag +J_{2k}a_-^\dag \right) + \beta_{s}\left(J^{\dagger}_{0k} a_+^\dag + J^{\dagger}_{-2k}a_-^\dag \right) + {\rm H.c.}\right\} .\label{12.13}
\end{eqnarray}
The parameters of the Hamiltonian are a function of the detunings and Rabi frequencies of the two highly detuned laser fields co-propagating with the clockwise cavity mode and thus could be controlled through the laser frequencies and intensities. 

Finally, we reformulate the Hamiltonian (\ref{12.13}) in terms of bosonic variables by adopting the Holstein-Primakoff representation of angular momentum operators~\cite{hp40}. In this representation, the collective atomic operators, $J^{\dagger}_{mk}, J_{mk}$ and $J_{z}$ are expressed in terms of annihilation and creation operators $C_{mk}$ and $C^{\dagger}_{mk}$ of a single bosonic mode
\begin{eqnarray}
J_{mk} = C_{mk}\sqrt{N -C_{mk}^\dag C_{mk}} ,\quad J_{z}=C_{mk}^\dag C_{mk} -N/2 .\label{e6u}
\end{eqnarray}
Provided the atoms in each ensemble are initially prepared in their ground states $|0_{j}\rangle$, and taking into account that due to large detunings of the driving fields, the excitation probability of each atom is low during the laser-atom-cavity coupling, i.e., the number of atoms transferred to the states $|1_{j}\rangle$ is expected to be much smaller than the total number of atoms in each ensemble, i.e., $\langle C_{mk}^\dag C_{mk}\rangle \ll N$. By expanding the square root in Eq.~(\ref{e6u}) and neglecting terms of the order of $\emph{O}(1/N)$, the collective atomic operators can be approximated~as
\begin{eqnarray}
J_{mk} = \sqrt{N} C_{mk} ,\quad J_{z} =  -N/2 ,\label{12.14}
\end{eqnarray}
where
\begin{eqnarray}
C_{mk} = \frac{1}{\sqrt{N}}\sum\limits_{j=1}^N b_{j}{\rm e}^{ imkx_{j}} ,\quad m=0,\pm 2 ,\label{12.15}
\end{eqnarray}
are collective bosonic operators with the operators $b_{j}$ and~$b^{\dagger}_{j}$ obeying the standard bosonic commutation relation $[b_{j}, b^{\dagger}_{\ell}]=\delta_{j\ell}$.

It is easy to prove that in the limit of $N\gg 1$,we obtain the standard bosonic commutation relation
\begin{eqnarray}
\left[C_{mk},C^{\dag}_{m^\prime k}\right] \approx \delta_{m,m^\prime} .\label{12.18}
\end{eqnarray}

Thus, in terms of the collective bosonic operators $C_{0k}$ and~$C_{\pm 2k}$, the effective Hamiltonian~(\ref{12.13}) takes the form
\begin{eqnarray}
H_{e} =  \hbar\left( \beta_{u}C_{0k}+\beta_{s}C_{0k}^{\dag}\right)a_+^\dag
+ \hbar\left( \beta_{u}C_{2k} + \beta_{s}C^{\dag}_{-2k}\right) a_-^\dag + {\rm H.c.}  \label{12.19}
\end{eqnarray}
The important property of the bosonic representation is the fact that the Hamiltonian of an ensemble of atoms trapped inside a ring cavity can be expressed as the interaction between the cavity modes and three orthogonal field modes. 

Since our objective is to prepare the atomic ensemble in a desired entangled state in the presence of a possible loss of photons due to the damping of the cavity mode, we shall work with the density operator $\rho$ of the system whose the evolution is governed by the master equation
\begin{eqnarray}
\dot{\rho} = -\frac{i}{\hbar}[H_{e} ,\rho]+{\cal L}_c\rho ,\label{12.21}
\end{eqnarray}
where
\begin{equation}
{\cal L}_{c}\rho = \frac{1}{2}\kappa \sum\limits_{i=\pm}\left(2a_{i}\rho
a^\dag_{i}-a^\dag_{i} a_{i}\rho-\rho a^\dag_{i} a_{i}\right) ,\label{12.22}
\end{equation}
is the Liuivilian operator representing the damping of the cavity field modes with the rate~$\kappa$.

In what follows, we show that by a proper choosing of the Rabi frequencies $\Omega_{u}, \Omega_{s}$ of the laser fields and their phases, $\phi_{u}, \phi_{s}$ one may prepare the system in a desired state that then decays with the rate $\kappa$ to a steady-state form.

\subsection{One and two-mode entangled states}

Let us now illustrate in details how to construct entangled states in a single ensemble of cold atoms located inside a two-mode ring cavity. We focus on the creation of single and two-mode entangled states and attempt to characterize the entanglement in terms of unitary operators known as squeezed operators. The operators for single and two-mode squeezed states are defined as~\cite{cs85,fd04}
\begin{eqnarray}
S_0(\xi_0) &=& \exp\left[-\frac{1}{2}\left(\xi_0C_{0k}^{\dag 2}-\xi_{0}^{\ast}C_{0k}^2\right)\right] ,\nonumber\\
S_{\pm k}(\xi_1) &=& \exp\left(\xi_{1}^{\ast} C_{2k}C_{-2k} - \xi_{1}C_{2k}^\dag C_{-2k}^\dag\right) ,\label{e21z}
\end{eqnarray}
where $\xi_{0}$ and $\xi_{1}$ are complex one and two-mode squeezing parameters, respectively.

Entangled (squeezed) states in a single ensemble are constructed from the vacuum by a unitary transformation associated with the realistic dynamical process determined by the master equation (\ref{12.19}). The creation of entangled states is done in two steps. In the first step, we adjust the driving lasers to propagate in the clockwise direction, along the cavity mode $(+)$. In this case, the dynamics of the system are determined by the Hamiltonian~(\ref{12.19}). We then send series of laser pulses of phases $\phi_{u}=\phi_{s}=0$ and arbitrary Rabi frequencies~$\Omega_{u}$ and $\Omega_{s}$, but such that $|\beta_{u}| > |\beta_{s}|$. With this choice of the parameters of the driving lasers and under the unitary squeezing transformation
\begin{eqnarray}
S_0(-\xi_{0})S_{\pm k}(-\xi_{1})\rho S_0(\xi_0)S_{\pm k}(\xi_{1}) = \tilde{\rho} ,\label{12.24}
\end{eqnarray}
with
\begin{eqnarray}
\xi_0=\xi_{1}=\frac{1}{2}\ln\left(\frac{|\beta_{u}|+|\beta_{s}|}{|\beta_{u}| - |\beta_{s}|}\right) ,\label{12.25}
\end{eqnarray}
the master equation (\ref{12.21}) becomes
\begin{equation}
\frac{d}{dt}\tilde{\rho}=-i[\tilde{H_e},\tilde{\rho}]+ {\cal L}_{c}\tilde{\rho} ,\label{12.26}
\end{equation}
where
\begin{eqnarray}
\tilde{H_e} &=& S_0(-\xi_0)S_{\pm k}(-\xi_{1})H_eS_0(\xi_0)S_{\pm k}(\xi_{1})\nonumber\\
&=&\hbar\sqrt{|\beta_{u}|^2 - |\beta_{s}|^2} \left(a_+^\dag C_{0k} + a_-^\dag C_{2k} + {\rm H.c.}\right) .\label{12.27}
\end{eqnarray}
It is seen that under the squeezing transformation, the Hamiltonian represents a simple system of two independent linear mixers, where the collective bosonic modes~$C_{0k}$ and $C_{2k}$ linearly couple to the cavity modes~$a_{+}$ and~$a_{-}$, respectively. The mode $C_{-2k}$ is decoupled from the cavity modes and therefore does not evolve. In other words, the state of the mode~$C_{-2k}$ cannot be determined by the evolution operator for the Hamiltonian~(\ref{12.27}). The important property of the transformed system is that the master equation~(\ref{12.26}) is fully soluble, i.e. all eigenvectors and eigenvalues can be obtained exactly. Hence, we can monitor the evolution of the bosonic modes towards their steady-state values. Since we are interested in the steady-state of the system, we confine our attention only to the eigenvalues of Eq.~(\ref{12.26}), which are of the form
\begin{eqnarray}
\eta_\pm = -\frac{\kappa}{2}\pm\left[\left(\frac{\kappa}{2}\right)^2-\sqrt{|\beta_{u}|^{2}-|\beta_{s}|^{2}}\right]^\frac{1}{2} .\label{12.28}
\end{eqnarray}
Evidently, both eigenvalues have negative real parts which means that the system subjected to a series of laser pulses up to a short time $t$ will evolve (decay) to a stationary state that is a vacuum state. Thus, as a result of the interaction given by the Hamiltonian (\ref{12.27}), and after a sufficiently long evolution time, the modes $a_{\pm}$, $C_{0k}$ and $C_{2k}$ will be found in the vacuum state, whereas the mode $C_{-2k}$ will remain in an undetermined state. The state of the mode $C_{-2k}$ will be determined in the next, second step of the preparation process.

In order to estimate the time scale for the system to reach the steady-state, we see from~Eq.~(\ref{12.28}) that as long as $\sqrt{|\beta_{u}|^{2}-|\beta_{s}|^{2}}>\kappa/2$, the time scale for the system  to reach the steady state is of order of $\simeq 2/\kappa$. Thus, as a result of the cavity damping the system, after a sufficient long time, will definitely be found in the stationary state.

Thus, after the first step of the preparation, we find that in the steady-state the density matrix representing the state of the transformed system is in the factorized form
\begin{eqnarray}
\tilde{\rho}(\tau\simeq 2/\kappa)= \tilde{\rho}_{v}\otimes\tilde{\rho}_{C_{-2k}} ,\label{12.29}
\end{eqnarray}
where
\begin{eqnarray}
\tilde{\rho}_{v} =|0_{a_+},0_{a_-},0_{C_{0k}},0_{C_{2k}}\rangle
\langle 0_{a_+},0_{a_-},0_{C_{0k}},0_{C_{2k}}|  \label{12.30}
\end{eqnarray}
is the density matrix of the four modes prepared in their vacuum states, and $\tilde{\rho}_{C_{-2k}}$ is the density matrix of the mode $C_{-2k}$ whose the state has not yet been determined. The ket $|0_{a_+},0_{a_-},0_{C_{0k}},0_{C_{2k}}\rangle$ represents the state with zero photons in each of the modes.

What left is to prepare the remaining collective mode $C_{-2k}$ in a desired squeezed vacuum state. This is done in what we call the second step of the preparation, in which we first adjust the driving lasers to propagate along the anti-clockwise mode $(-)$. We then send series of pulses of frequencies, phases and amplitudes the same as in the above first step. As a result of the coupling to the cavity mode~$(-)$,  the interaction is now governed by the Hamiltonian (\ref{12.22}), and therefore after the unitary squeezing transformation the Hamiltonian of the system takes the form
\begin{eqnarray}
\tilde{H_e} &=& S_0(-\xi_0)S_{\pm k}(-\xi_{1})H_eS_0(\xi_0)S_{\pm k}(\xi_{1})\nonumber\\
&=&\hbar \sqrt{|\beta_{u}|^2 - |\beta_{s}|^2} \left( a_-^\dag C_{0k} + a_+^\dag C_{-2k} + {\rm H.c.}\right) .\label{12.31}
\end{eqnarray}
As above in the case of the coupling to the cavity mode~$a_{+}$, the Hamiltonian (\ref{12.31}) describes a system of two independent linear mixers. Hence, the state of the system will evolve during the interaction   towards its stationary value, and after a suitably long time,~$\simeq 2/\kappa$, the transformed system will be found in the vacuum state.

Hence, after the second step of the preparation, the transformed system is found in the pure vacuum state determined by the density matrix of the form
\begin{equation}
\tilde{\rho}(\tau\simeq 4/\kappa)=|\tilde{\Psi}\rangle\langle\tilde{\Psi} | ,\label{12.32}
\end{equation}
where
\begin{eqnarray}
|\tilde{\Psi}\rangle = S_0(-\xi_0)S_{\pm k}(-\xi_{1})|\Psi\rangle
= |0_{a_{+}},0_{a_{-}},0_{C_{0k}},0_{C_{2k}},0_{C_{-2k}}\rangle \label{12.33}
\end{eqnarray}
represents the vacuum state of the transformed system and the ket $|\Psi\rangle$ represents the final stationary state of the system.

If we now perform the inverse transformation from~$|\tilde{\Psi}\rangle$ to $|\Psi\rangle$, we find that the system is in the multi-mode pure squeezed state
\begin{eqnarray}
|\Psi\rangle = S_0(\xi_0)S_{\pm k}(\xi_{1})|0_{C_{0k}},0_{C_{2k}},0_{C_{-2k}}\rangle\otimes |0_{a_{+}},0_{a_{-}}\rangle .\label{12.34}
\end{eqnarray}
The density operator representing the multi-mode squeezed state (\ref{12.34}) is
\begin{eqnarray}
\rho = S_0(\xi_0)S_{\pm k}(\xi_{1})|\{0\}\rangle\langle \{0\}|S_0(-\xi_0)S_{\pm k}(-\xi_{1}) ,\label{12.34a}
\end{eqnarray}
where $|\{0\}\rangle =|0_{C_{0k}},0_{C_{2k}},0_{C_{-2k}}\rangle\otimes|0_{a_{+}},0_{a_{-}}\rangle$.

Equations (\ref{12.34}) and (\ref{12.34a}) show that in the steady-state the cavity modes are left in the vacuum state and the atomic ensemble is prepared in single and two-mode squeezed states. In other words, it shows that the collective mode $C_{0k}$ is prepared in the one-mode squeezed vacuum state $S_0(\xi_0)|0_{C_{0k}}\rangle$, whereas the collective modes~$C_{\pm2k}$ are in the two-mode squeezed vacuum state $S_{\pm k}(\xi_{1})|0_{C_{2k}},0_{C_{-2k}}\rangle$ associated with the superposition of two, position dependent counter-propagating modes. The superposition nature of the squeezed states allows us to conclude that the atomic ensemble, after the interaction with the sequences of the laser pulses is prepared in single and two-mode entangled states.

\section{Conclusions}

We have considered to examples of application of the CV approach to entanglement creation and processing. In the first, we have analyzed the time evolution of an initial entanglement encoded into two coupled harmonic oscillators interacting with a thermal reservoir. We have shown that it is very convenient to analyze dynamics of the oscillators in terms of the evolution of the covariance matrix elements. By performing a suitable transformation of the position and momentum operators of the system oscillators, we have shown that the set of ten coupled differential equations for the covariance matrix elements splits into decoupled subsets of smaller sizes involving only three and four equations. A general feature of the entanglement evolution is that it exhibits two characteristic time scales, a shot time regime where an initial entanglement is rapidly damped and a long time regime where the entanglement undergoes continuous undamped oscillations. Depending on the initial amount of entanglement encoded into the system, it can be preserved for all times or may undergo the sudden death and revival phenomena. We have also found that in contrast to what one could expect, a stronger coupling of the oscillators to the thermal reservoir leads to a better stationary entanglement than in the case of a weak coupling.

In the second example of application of the CV approach, we have described a practical scheme for the creation of CV entangled states of effective bosonic modes realized in an atomic ensemble interacting collectively with two counter propagating modes of a ring cavity. The basic idea of the scheme is to transfer the ensemble field modes into suitable linear combinations that can be prepared, by a sequential application of the laser pulses, in pure squeezed vacuum (entangled) states.

\Thebibliography

\bibitem{bp03} S. L. Braunstein and A. K. Pati, {\it Quantum Information Theory
with Continuous Variables}, Kluwer, Dordrecht, 2003.

\bibitem{s1} S. L. Braunstein and P. van Loock, Quantum information with continuous variables, {\it Rev. Mod. Phys.} {\bf 77} (2005), 513-577.

\bibitem{s2} A. Furusawa, J. L. Sorensen, and S. L. Braunstein, C. A. Fuchs, and H. J. Kimble and E. S. Polzik, Unconditional quantum teleportation, {\it Science} {\bf 282} (1998), 706-709; N. Takei, T. Aoki, S. Koike, K. Yoshino, K. Wakui, H. Yonezawa, T. Hiraoka, J. Mizuno, M. Takeoka, M. Ban, and A. Furusawa, Experimental demonstration of quantum teleportation of a squeezed state, {\it Phys. Rev. A} {\bf 72} (2005), 042304; H. Yonezawa, S. L. Braunstein, and A. Furusawa,  Experimental demonstration of quantum teleportation of broadband squeezing, {\it Phys. Rev. Lett.} {\bf 99} (2007), 110503.

\bibitem{s3} X. Y. Li, Q. Pan, J. T. Jing, J. Zhang, C. D. Xie, and K. Peng, Quantum dense coding exploiting a bright Einstein-Podolsky-Rosen beam, {\it Phys. Rev. Lett.} {\bf 88} (2002), 047904.

\bibitem{s4} X. J. Jia, X. L. Su, Q. Pan, J. R. Gao, C. D. Xie, and K. Peng, Experimental demonstration of unconditional entanglement swapping for continuous variables, {\it Phys. Rev. Lett.} {\bf 93} (2004), 250503.

\bibitem{s5} N. Takei, H. Yonezawa, T. Aoki, and A. Furusawa, High-fidelity teleportation beyond the no-cloning limit and entanglement swapping for continuous variables, {\it Phys. Rev. Lett.} {\bf 94} (2005), 220502; J. Zhang, C. Xie, K. Peng, and P. van Loock, Anyon statistics with continuous variables, {\it Phys. Rev. A} {\bf 78} (2008), 052121.

\bibitem{lg07} K.-L. Liu and H.-S. Goan, Non-Markovian entanglement dynamics of quantum continuous variable systems in thermal environments, {\it Phys. Rev. A} {\bf 76} (2007), 022312.

\bibitem{paz92} B. L. Hu, J. P. Paz and Y. Zhang, Quantum Browian-motion in a general environment - exact master equation with nonlocal dissipation and colored noise, {\it Phys. Rev. D} {\bf 45} (1992), 2843-2861.

\bibitem{mo07} S. Maniscalco, S. Olivares, and M. G. A. Paris,  Entanglement oscillations in non-Markovian quantum channels, {\it Phys. Rev. A} {\bf 75} (2007), 062119.

\bibitem{hb08} C. H\"orhammer and H. B\"uttner, Environment-induced two-mode entanglement in quantum Brownian motion, {\it Phys. Rev. A} {\bf 77} (2008), 042305.

\bibitem{b02} D. Braun, Creation of entanglement by interaction with a common heat bath, {\it Phys. Rev. Lett.} {\bf 89} (2002), 277901.

\bibitem{kl02} M. S. Kim, J. Lee, D. Ahn, and P. L. Knight, Entanglement induced by a single-mode heat environment, {\it Phys. Rev. A} {\bf 65} (2002), 040101.

\bibitem{bf03} F. Benatti, R. Floreanini, and M. Piani, Environment induced entanglement in Markovian dissipative dynamics, {\it Phys. Rev. Lett.} {\bf 91} (2003), 70402.

\bibitem{ok06} S. Oh and J. Kim, Entanglement between qubits induced by a common environment with a gap, {\it Phys. Rev. A} {\bf 73} (2006), 062306.

\bibitem{bl07} B. Bellomo, R. Lo Franco, and G. Compagno, Non-Markovian effects on the dynamics of entanglement, {\it Phys. Rev. Lett.} {\bf 99} (2007), 160502; Entanglement dynamics of two independent qubits in environments with and without memory, {\it Phys. Rev. A} {\bf 77} (2008), 032342.

\bibitem{p02} M. G. A. Paris, Optimized quantum nondemolition measurement of a field quadrature, {\it Phys. Rev. A} {\bf 65} (2002), 012110.

\bibitem{sl04} A. Serafini, F. Illuminati, M. G. A. Paris, and S. De Siena, Entanglement and purity of two-mode Gaussian states in noisy channels, {\it Phys. Rev. A} {\bf 69} (2004), 022318.

\bibitem{dh04} P. J. Dodd and J. J. Halliwell, Disentanglement and decoherence by open system dynamics, {\it Phys. Rev. A} {\bf 69} (2004), 052105.

\bibitem{pb04} J. S. Prauzner-Bechcicki, Two-mode squeezed vacuum state coupled to the
common thermal reservoir, {\it J. Phys. A: Math. Gen.} {\bf 37} (2004), L173-L181.

\bibitem{bf06} F. Benatti and R. Floreanini, Entangling oscillators through environment noise, {\it J. Phys. A: Math. Gen.} {\bf 39} (2006), 2689-2699.

\bibitem{az07} J.-H. An and W.-M. Zhang, Non-Markovian entanglement dynamics of noisy continuous-variable quantum channels, {\it Phys. Rev. A} {\bf 76} (2007), 042127.

\bibitem{cb08} C.-H. Chou, T. Yu, and B. L. Hu, Exact master equation and quantum decoherence of two coupled harmonic oscillators in a general environment, {\it Phys. Rev. E} {\bf 77} (2008), 011112.

\bibitem{pa09} M. Paternostro, G. Adesso, and S. Campbell, Passing quantum correlations to qubits using any two-mode state, {\it Phys. Rev. A} {\bf 80} (2009), 062318.

\bibitem{pr08} J. P. Paz and A. J. Roncaglia, Dynamics of the entanglement between two oscillators in the same environment, {\it Phys. Rev. Lett.} {\bf 100} (2008), 220401; Dynamical phases for the evolution of the entanglement between two oscillators coupled to the same environment, {\it Phys. Rev. A} {\bf 79} (2009), 032102.

\bibitem{Yu} T. Yu and J. H. Eberly, Finite-time disentanglement via spontaneous emission, {\it Phys. Rev. Lett.} \textbf{93} (2004), 140404.

\bibitem{ey07} T. Yu and J. H. Eberly, Sudden Death of Entanglement, {\it Science} \textbf{323} (2009), 598-601.

\bibitem{ft06} Z. Ficek and R. Tana\'s,  Dark periods and revivals of entanglement in a two-qubit system, {\it Phys. Rev. A} \textbf{74} (2006), 024304.

\bibitem{cz08} X. Cao and H. Zheng, Non-Markovian disentanglement dynamics of a two-qubit system, {\it Phys. Rev. A} {\bf 77} (2008), 022320.

\bibitem{mm09} L. Mazzola, S. Maniscalco, J. Piilo, K.-A. Suominen, and B. M. Garraway, Sudden death and sudden birth of entanglement in common structured reservoirs, {\it Phys. Rev. A} {\bf 79} (2009), 042302.

\bibitem{jl09} J. Jing, Z.-G. L\"u, and Z. Ficek, Breakdown of the rotating-wave approximation in the description of entanglement of spin-anticorrelated states, {\it Phys. Rev.  A} {\bf 79} (2009), 044305.

\bibitem{s12} M. D. Lukin, Colloquium: Trapping and manipulating photon states in atomic ensembles, {\it Rev. Mod. Phys.} {\bf 75} (2003), 457-472.

\bibitem{s13} M. Fleischhauer, A. Imamoglu, and J. P. Marangos, Electromagnetically induced transparency: Optics in coherent media, {\it Rev. Mod. Phys.} {\bf 77} (2005), 663-673.

\bibitem{s14} K. Hammerer, A. S. Sorensen, and E. S. Polzik, Quantum interface between light and atomic ensembles, {\it Rev. Mod. Phys.} {\bf 82} (2010), 1041-1093.

\bibitem{lf10} G.-X. Li and Z. Ficek, Creation of pure multi-mode entangled states in a ring cavity, {\it Optics Commun.} {\bf 283} (2010), 814-821.

\bibitem{s15} D. N. Matsukevich, T. Chaneliere, S. D. Jenkins, S. Y. Lan, T. A. B. Kennedy, and A. Kuzmich, Deterministic single photons via conditional quantum evolution, {\it Phys. Rev. Lett.} {\bf 97} (2006), 013601; J. Simon, H. Tanji, J. K. Thompson, and V. Vuletic, Interfacing collective atomic excitations and single photons, {\it Phys. Rev. Lett.} {\bf 98} (2007), 183601; C. W. Chou, J. Laurat, H. Deng, K. S. Choi, H. de Riedmatten, D. Felinto, and H. J. Kimble, Functional quantum nodes for entanglement distribution over scalable quantum networks, {\it Science} {\bf 316} (2007), 1316-1320; Z. S. Yuan, Y. A. Chen, B. Zhao, S. Chen, J. Schmiedmayer, and J. W. Pan, Experimental demonstration of a BDCZ quantum repeater node, {\it Nature} {\bf 454} (2008), 1098-1101.

\bibitem{s19} A. Chia  and A. S. Parkins, Entangled-state cycles of atomic collective-spin states, {\it Phys. Rev. A} {\bf 77} (2008), 033810.

\bibitem{s20} A. S. Parkins, E. Solano, and J. I. Cirac, Unconditional two-mode squeezing of separated atomic ensembles, {\it Phys. Rev. Lett.} {\bf 96} (2006), 053602.

\bibitem{lk09} G. X. Li, S. S. Ke, and Z. Ficek, : Generation of pure continuous-variable entangled cluster states of four separate atomic ensembles in a ring cavity, {\it Phys. Rev. A} {\bf 79} (2009), 033827.

\bibitem{s11} M. Yukawa, R. Ukai, P. van Loock, and A. Furusawa,  Experimental generation of four-mode continuous-variable cluster states, {\it Phys. Rev. A} {\bf 78} (2008), 012301.

\bibitem{l06} G.-X. Li, Generation of pure multipartite entangled vibrational states for ions trapped in a cavity, {\it Phys. Rev. A} {\bf 74} (2006), 055801.

\bibitem{dic} R. H. Dicke, Coherence in spontaneous radiation processes, {\it Phys. Rev.} {\bf 93} (1954), 99-110.

\bibitem{ft02} Z. Ficek and R. Tana\'s,  Entangled states and collective nonclassical effects in two-atom systems, {\it Phys. Rep.} {\bf 372} (2002), 369-443.

\bibitem{Pathintegralapproach_Caldeira1983} A. O. Caldeira and A. J. Leggett, Path integral approach to quantum Browian-motion, {\it Physica A} {\bf 121} (1983), 587-616.

\bibitem{Atom_Photon_Cohen92} C.~Cohen-Tannoudj, J.~Dupont-Roc, and G.~Grynberg, {\it Atom-Photon Interactions}, Wiley, New York, 1992.

\bibitem{dg00} L.-M. Duan, G. Giedke, J. I. Cirac, and P. Zoller, Inseparability criterion for continuous variable systems, {\it Phys. Rev. Lett.} {\bf 84} (2000), 2722-2725.

\bibitem{SeparabilityCriterion_Simon00} R.~Simon, Peres-Horodecki separability criterion for continuous variable systems, {\it Phys. Rev. Lett.} {\bf 84} (2000), 2726-2729.

\bibitem{de07} F. Dimer, B. Estienne, A. S. Parkins, and H. J. Carmichael, Proposed realization of the Dicke-model quantum phase transition in an optical cavity QED system, {\it Phys. Rev. A} {\bf 75} (2007), 013804.

\bibitem{cb09} M. M. Cola, D. Bigerni, and N. Piovella, Recoil-induced subradiance in an ultracold atomic gas, {\it Phys. Rev. A} {\bf 79} (2009), 053622.

\bibitem{li06} G. X. Li, S. P. Wu, and G. M. Huang, Generation of entanglement and squeezing in the system of two ions trapped in a cavity, {\it Phys. Rev. A} {\bf 71} (2005), 063817.

\bibitem{krb03} D. Kruse, M. Ruder, J. Benhelm, C. von Cube, C. Zimmermann, Ph. W. Courteille, Th. Els\"asser, B. Nagorny, and A. Hemmerich, Cold atoms in a high-Q ring cavity, {\it Phys. Rev. A} {\bf 67} (2003), 051802.

\bibitem{na03} B. Nagorny, Th. Els\"asser, H. Richter, A. Hemmerich, D. Kruse, C. Zimmermann, and Ph. Courteille, Optical lattice in a high-finesse ring resonator, {\it Phys. Rev. A} {\bf 67} (2003), 031401.

\bibitem{kl06} J. Klinner, M. Lindholdt, B. Nagorny, and A. Hemmerich, Normal mode splitting and mechanical effects of an optical lattice in a ring cavity, {\it Phys. Rev. Lett.} {\bf 96} (2006), 023002.

\bibitem{hp40} T. Holstein and H. Primakoff, Field dependence of the intrinsic domain magnetization of a ferromagnet, {\it Phys. Rev.}  {\bf 58} (1940), 1098-1113.

\bibitem{cs85} C. M. Caves and B. L. Schumaker, New formalism for 2-photon quantum optics: 1. Quadrature phases and squeezed states, {\it Phys. Rev. A}  {\bf 31} (1985), 3068-3092.

\bibitem{fd04} {\it Quantum Squeezing}, edited by P. D. Drummond and Z. Ficek, Springer, New York, 2004.

\end{thebibliography}

\end{document}